\documentclass{style/nseJournal}

\usepackage{booktabs} 

\begin{document}

\title{Interplay of Variance Reduction and Population Control \\ in Monte Carlo Neutron Transport} 

\addAuthor{\correspondingAuthor{Jordan Northrop}}{a}
\correspondingEmail{northroj@oregonstate.edu}
\addAuthor{Ilham Variansyah}{a}
\addAuthor{Todd Palmer}{a}
\addAuthor{Camille J. Palmer}{a}

\addAffiliation{a}{School of Nuclear Science and Engineering, Oregon State University}

\addKeyword{Monte Carlo}
\addKeyword{Radiation Transport}
\addKeyword{Variance Reduction}
\addKeyword{Population Control}
\addKeyword{Exascale}
 
\titlePage

\begin{abstract}

Monte Carlo methods are widely used for neutron transport simulations at least partly because of the accuracy they bring to the modeling of these problems.
However, the computational burden associated with the slow convergence rate of Monte Carlo poses a significant challenge to running large-scale simulations.
The continued improvement in high-performance computing capabilities has put exascale time-dependent Monte Carlo neutron transport simulations within reach.
Variance reduction techniques have become an essential component to the efficiency of steady-state simulations, and population control techniques are an integral part of time-dependent simulations, but combining them can create algorithmic conflicts.
This study investigates the performance of steady-state variance reduction techniques when extended to time-dependent problems and examines how variance reduction and population control techniques combine to impact the effectiveness of time-dependent simulations.
Simulations were conducted using various combinations of these techniques across multiple test problems to assess their performance.
While this study does not examine all possible variance reduction and population control combinations, the findings emphasize the importance of carefully selecting algorithms to simulate large-scale time-dependent problems effectively.
Notably, using weight windows with weight-based combing for population control can significantly hinder simulation performance, whereas pairing weight windows with uniform combing can provide the efficiencies necessary for successfully computing the results of massive problems.
Further performance gains were observed when steady-state weight windows were replaced with time-dependent versions.

\end{abstract}

\section*{List of Acronyms}
\begin{tabular}{ll}
VRT  & Variance Reduction Technique \\
IC & Implicit Capture \\
CADIS & Consistent Adjoint Driven Importance Sampling\\
FWCADIS & Forward Weighted Consistent Adjoint Driven Importance Sampling\\
CL & Cooper-Larsen Weight Window Method\\
PCT  & Population Control Technique \\
UC & Uniform Combing \\
WC & Weight-Based Combing \\
FOM & Figure of Merit \\
JIT & Just-In-Time Compilation \\
CSG & Constructive Solid Geometry
\end{tabular}

\section{Introduction}

Monte Carlo methods are widely recognized for accurately producing results in neutron transport simulations, minimizing the need for discretization errors by treating each dimension of phase space continuously. 
The major limitation of Monte Carlo is its slow convergence rate, proportional to the square root of the number of particle histories.
This can make achieving tallies with low statistical noise computationally expensive for complex problems. 
To address this, a variety of variance reduction techniques (VRTs) have been developed and are widely used to boost the efficiency of calculations \cite{Wagner1998,Wagner2014, Davis2011, Valentine2022}.

The high computational demand historically limited Monte Carlo applications to steady-state or eigenvalue simulations.
With recent advancements in computational power, fully time-dependent Monte Carlo neutron transport simulations of large scale problems are becoming more practical ~\cite{Faucher2018, Ferraro2020}. 
Improvements to time-dependent Monte Carlo algorithms have also contributed to this increase in capability ~\cite{Sjenitzer2013, Faucher2018, Valentine2022, Montecchio2022}.
The additional computational burden of time dependence can increase the value of efficient VRTs, but most established VRTs were designed and validated in steady-state contexts. 
Furthermore, many VRTs rely on modifying particle weights, which can conflict with popular population control techniques (PCTs) that are crucial for time-dependent simulations \cite{Sjenitzer2013, Variansyah2022}.

Other works have investigated time-dependent VRTs and their interactions with time-dependent algorithms like PCTs ~\cite{Legrady2020, Frohlicher2023, Montecchio2022}, but the focus of this work is on global VRTs that modify the spatial (or other non-temporal dimensions) distribution of particles and their weights.
These methods have primarily been developed and applied to steady-state shielding problems ~\cite{Wagner2014, Cooper2001}.
This work demonstrates the application of time-independent global VRTs to time-dependent simulations, explores the interactions between certain common weight-based global VRTs and PCTs, and identifies specific adaptations of VRTs for time-dependence that can further improve simulation efficiency.

\section{Variance Reduction}

Analog Monte Carlo neutron transport simulations closely replicate physical processes but often fail to efficiently direct particles to areas of the problem domain where their contribution is most needed for statistical resolution \cite{MCNPmanual}. 
To address this, variance reduction techniques modify particle weights based on their distribution in the phase space, guiding particles more effectively to regions of interest. 
Importantly, weight adjustments must be carefully managed to preserve the expected values of quantities of interest.

Two primary VRT operations, splitting and rouletting, are widely used to modify particle weights and control particle counts to improve statistical efficiency.
Splitting involves duplicating a particle into several particles, each with a reduced weight such that the total weight remains unchanged. 
This allows more particles to be directed to areas where increased particle counts enhance the precision of tallies. 
Conversely, rouletting reduces particle counts by probabilistically terminating particles based on a probability $p$. 
If a particle survives this process, its weight is adjusted to $\frac{1}{1-p}$ to conserve the expected tally. 
Rouletting thus minimizes the computational load in areas where particles have minimal impact on the final results.

The specific types of VRTs discussed here are common in steady-state calculations, particularly shielding calculations.
Many other types of VRTs exist and have been developed/tested with time-dependence and PCTs, but those methods are not the focus of this work ~\cite{Legrady2020, Frohlicher2023}.

\subsection{Implicit Capture}

Implicit capture (IC), also known as survival biasing, is one of the simplest and most widely used variance reduction techniques \cite{MCNPmanual}. 
IC enhances particle movement through absorbing regions by effectively removing absorption from the list of potential collision outcomes. 
Instead of being absorbed, the particle continues with a reduced weight, selecting an alternative reaction path that keeps it in transit.
The particle’s weight is set to $w_{out} = w_{in}*(1 - p_a)$, where $p_a$ represents the probability of absorption. 
To prevent particles from undergoing excessively prolonged histories when IC is applied, Monte Carlo codes often enforce a lower weight threshold using rouletting.

\subsection{Weight Windows}

Weight windows provide a versatile approach to control particle splitting and rouletting within specific areas of a simulation. 
Typically, a spatial or temporal mesh is applied, dividing the domain into cells, each with a central weight that defines an acceptable range of particle weights. 
This range is determined by scaling the central weight to set a ceiling and a floor. 
Particles with weights exceeding the ceiling are split to bring their weights within the range, while those below the floor have their weights increased or are terminated through rouletting. 
Weight windows can be applied across all dimensions of the phase space, but in this paper, only spatial and temporal meshes are considered.

For localized variance reduction, such as in source-detector problems where accurate results are needed at particular locations, techniques like Consistent Adjoint-Driven Importance Sampling (CADIS)~ \cite{Wagner1998} are highly effective. 
CADIS uses an adjoint simulation to generate an importance map that biases particles toward the tally of interest.
The adjoint flux from this simulation indicates where particles contribute most to the tally. 
Areas of high adjoint flux show where particles should be retained because they are more likely to contribute to the tally, while low adjoint flux regions suggest where particle counts can be reduced.
 
CADIS operates in two stages.
First, a deterministic radiation transport code generates the adjoint flux, with the tally of interest serving as the adjoint source. 
Then, the Monte Carlo code conducts the full forward simulation, with weight window centers defined as inversely proportional to the adjoint flux distribution. 
The window center $w$ in each mesh cell, given by the adjoint flux $\phi^\dagger$ and a normalization factor $R$, is:
\begin{equation} \label{eq:cadis1}
  w(\vec{r}, E,\vec{\Omega}) = \frac{R}{\phi^{\dagger}(\vec{r}, E, \vec{\Omega})}
\end{equation}
In addition, the forward source in the Monte Carlo simulation is biased to make the method \textit{consistent}.
The biased source is defined in Eq. (\ref{eq:cadis2}), where the new source $\hat{q}$, is calculated from the adjoint flux, the original source $q$, and a normalization factor.
\begin{equation} \label{eq:cadis2}
  \hat{q}(\vec{r},E, \vec{\Omega}) = \frac{\phi^{\dagger}(\vec{r}, E, \vec{\Omega})*q(\vec{r}, E, \vec{\Omega})}{R}.
\end{equation}
This biasing ensures particles are appropriately weighted from birth, aligning them with the local weight window rather than splitting or rouletting immediately upon creation, which reduces computational overhead.
CADIS involves the splitting of particles where the adjoint flux is high and the rouletting of particles where the adjoint flux is low.
In this way, computational effort is spent where the particles are likely to contribute to the tally and avoided where the particles are unlikely to impact the tally.

When high statistical accuracy is needed across an extensive portion of the problem domain, local variance reduction techniques become insufficient.
Global variance reduction requires optimizing computational resources across the entire domain, which poses significant challenges.
For global variance reduction, the optimal distribution of that variance across the problem is uniform.
Because the variance is proportional to the number of particle tracks contributing to the tallies, the following variance reduction techniques aim to distribute Monte Carlo particles as evenly as possible.

The Cooper-Larsen (CL) method addresses global variance reduction by leveraging the relationship between the scalar flux distribution $\phi$, particle distribution $m$, and weight distribution $w$ \cite{Cooper2001}:,
\begin{equation} \label{eq:cl1}
  \phi \propto m * w.
\end{equation}
To achieve a uniform particle distribution $m$, the weights $w$ should be proportional to the flux. 
However, the requirement to know the flux in advance creates a cyclical dependency. 
This challenge is typically resolved by performing an initial, low-cost simulation to approximate the flux.
A deterministic simulation is often used to generate this flux estimate, but a small-scale Monte Carlo run or mid-simulation weight adjustments can also be effective.~\cite{Zammataro2024}
While inaccuracies in the initial flux estimate may impact the efficiency of the weight windows, they should not affect the final solution’s accuracy in the full-scale simulation.

The Forward Weighted Consistent Adjoint Driven Importance Sampling (FW-CADIS) \cite{Wagner2014} method offers a flexible approach for reducing global variance. It combines the importance mapping of CADIS with the uniform particle distribution objective of the Cooper-Larsen method.

FW-CADIS aims to generate an adjoint map that indicates where particles contribute to achieving a uniform response across the entire problem domain. 
Unlike traditional CADIS, which defines a local tally as the adjoint source, FW-CADIS establishes the adjoint source over the full domain. 
The optimal adjoint source for achieving uniform particle density is defined as $q^\dagger(x) = 1/\phi(x)$ where $q^\dagger(x)$ represents the adjoint source and $\phi(x)$ is the scalar flux. \cite{Wagner2014}

The FW-CADIS procedure involves three main steps: first, a preliminary simulation provides an estimate of the forward flux; second, this flux is used to weight an adjoint source for an adjoint simulation; finally, the adjoint flux generated is applied to set weight windows for the primary Monte Carlo transport simulation (Eq. (\ref{eq:cadis1})) and apply source biasing (Eq. (\ref{eq:cadis2})). 
FW-CADIS provides more optimal weight windows for global variance reduction than CADIS, and it can also be adapted for targeted sections of the problem domain. 
For example, when multiple detectors surround an experiment, FW-CADIS can simultaneously optimize weight windows for multiple tallies.

\section{Population Control}

In time-dependent Monte Carlo neutron transport simulations, the particle population often fluctuates between time steps, leading to uneven variances across tally bins as some time steps receive more particle contributions than others. 
This imbalance affects statistical accuracy and presents practical challenges.
In supercritical systems, the rapid increase in particle count can make memory and computational demands unmanageable. 
Population control techniques, a subset of variance reduction techniques, address these issues by regulating the particle population over time, ensuring consistent and feasible simulation performance.

While there are various PCTs, this work focuses on two: uniform combing and weight-based combing \cite{Variansyah2022}. 
Both techniques are applied at the end of each time step, once all particles have been gathered in a census bank. 
At that point, the particle count is adjusted to a target level, and the census bank then serves as the source for the subsequent time step.

\begin{figure}
  \centering
  \includegraphics[trim = 10mm 0mm 10mm 0mm, width=140mm]{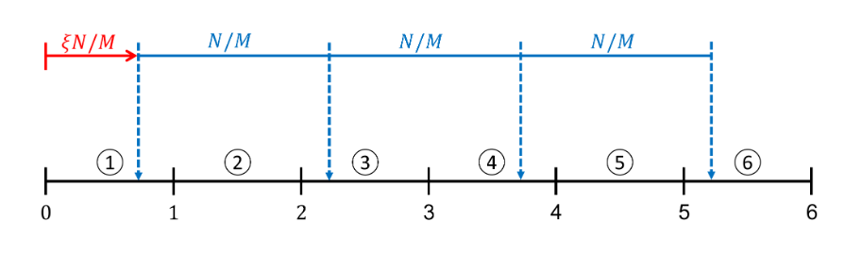}
  \caption{Depiction of uniform combing reducing $N=6$ particles to $M=4$ at census \cite{Variansyah2022}}
  \label{UCpicture}
\end{figure}

Uniform combing (UC) depicted in Figure \ref{UCpicture} treats all particles in the census bank equally, regardless of weight. 
Given $N$ particles at the end of a time step and a target of $M$ particles, UC lines up the particles on a number line from $1$ to $N$. 
A figurative comb with $M$ evenly spaced teeth, each separated by $N/M$, is then ``placed'' on the line with a random offset. 
Particles aligning with a tooth survive; otherwise, they are removed. 
Particles with multiple teeth pointing to them are duplicated. 
To preserve the overall solution, the total weight is conserved by scaling each surviving particle’s weight by a factor of $N/M$, shown in equation \ref{eq:uc1}.

\begin{equation} \label{eq:uc1}
  w_{final} = w_{initial} * N/M
\end{equation}

\begin{figure}
  \centering
  \includegraphics[trim = 10mm 0mm 10mm 0mm, width=140mm]{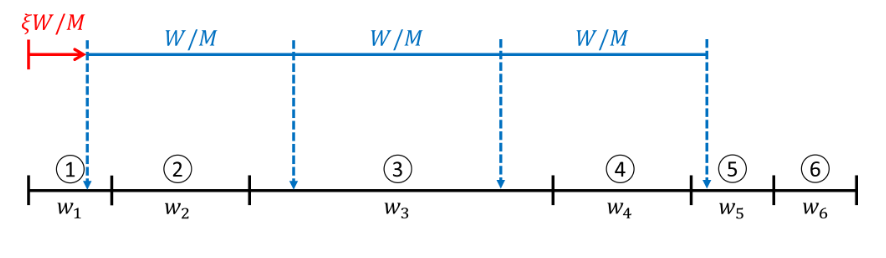}
  \caption{Depiction of weight-based combing reducing six particles to four at census \cite{Variansyah2022}}
  \label{WCpicture}
\end{figure}

Weight-based combing (WC) shown in Figure \ref{WCpicture} operates on a similar principle but assigns each particle a space on the number line proportional to its weight rather than a uniform length. 
The number line extends from $0$ to $W$, where $W$ represents the total weight in the census bank. 
A comb with $M$ teeth, spaced $W/M$ apart and incorporating a random offset, is then applied. 
This method increases the likelihood of larger-weight particles being selected, as their proportions on the number line are greater. 
Instead of scaling the weights of surviving particles by a constant factor, all surviving weights are reset to a uniform value of $W/M$.
This standardization of weights can reduce the variance because large differences in the scale of tallied quantities can lead to degraded variance.

\begin{equation} \label{eq:wc1}
  w_{final} = W/M
\end{equation}

Due to their interaction with particle weights, these PCTs may conflict with VRTs. 
Uniform combing is theoretically more compatible with VRTs since it does not alter the shape of the distribution of weights on average.
If a VRT has skewed the distribution of weights (and particles) in a specific manner, UC should primarily keep those distributions intact.
Only weight scaling and minor statistical artifacts from random removal/duplication of particles would need to be considered.
In contrast, weight-based combing significantly impacts both weight and particle distributions created through VRTs, leading to the hypothesis that it may coexist poorly with those methods ~\cite{Variansyah2022}.

\section{Methods}

Fully time-dependent Monte Carlo neutron transport codes are significantly less commonly available than steady-state/eigenvalue codes.
Time-dependent codes such as Mercury~\cite{Procassini2005}, Serpent~\cite{Leppanen2015}, and TRIPOLI-4~\cite{Brun2015}, have varying time-dependent features, but require specific licenses for use.
The most prevalent general purpose transport code, MCNP~\cite{MCNPmanual}, as well as the two main open source alternatives, OpenMC~\cite{Romano2015} and Geant4~\cite{Agostinelli2003}, do not yet have full time-dependent capabilties.
Many other simulation codebases exist, but are less commonly used.

Global VRTs are more challenging computationally than local VRTs, so problems that emphasize global variance reduction are a focus of our investigation ~\cite{Cooper2001}.
Two distinct approaches were utilized to evaluate VRTs and PCTs for time-dependent scenarios.

The first approach involves using an established code with existing advanced variance reduction capabilities and newly added time-dependent features. 
Shift, developed by Oak Ridge National Laboratory (ORNL), is a modern Monte Carlo code primarily focused on eigenvalue and steady-state problems \cite{Pandya2016}. 
Recently, a time-dependent branch has been introduced, which includes basic time-stepping and census capabilities~\cite{Reynolds2022}.
In this rudimentary time-dependent version of Shift, weight windows can only be set over the spatial dimensions, so time-dependent weight windows are not an option.
Weight-based combing is the only population control method available, and it cannot be turned off in time-dependent simulations.
In addition, certain physics, like delayed neutrons, have no time dependence, which limits the variety of time-dependent problems that can be accurately modeled.
Despite these limitations, Shift serves as a valuable testing tool due to its coupling with the deterministic transport code DENOVO, enabling efficient adjoint flux generation for CADIS-based methods.
DENOVO is packaged with Shift in the Omnibus framework and can easily execute the same problem from Shift's input file with little modification. \cite{Evans2010}

The second approach employs a fully time-dependent Monte Carlo code, MC/DC, developed by the Center for Exascale Monte Carlo Neutron Transport (CEMeNT) \cite{morgan2024}. 
It is an accelerated Python codebase designed to be performant, scalable, and portable.
Designed specifically for time-dependent transport studies, MC/DC offers a wider array of time-dependent capabilities than Shift, including delayed neutrons and moving geometries. 
However, MC/DC is not currently coupled with a deterministic code, so CADIS methods cannot access adjoint fluxes, and weight windows for the Cooper-Larsen method must be generated using a small-scale forward Monte Carlo simulation within MC/DC. 
This code uniquely supports both uniform and weight-based combing and can generate fully time-dependent weight windows, providing a dedicated platform for investigating VRTs and PCTs in dynamic transport scenarios.

The Lassen supercomputer at LLNL was used to measure performance on several test problems.
The simulations were performed in parallel on 40 Power9 IBM CPU cores on one node of Lassen.

\section{Results and Discussion}

To assess the effectiveness of variance reduction techniques, it is essential to consider both variance and runtime. 
A commonly used figure of merit (FOM) for such analyses is defined as the inverse of the runtime $T$, multiplied by the variance $\sigma^2$:
\begin{equation}
  FOM = \frac{1}{T*\sigma^2}
\end{equation}
Here, $T$ is proportional to the number of particle histories, while $\sigma^2$  is inversely proportional to the number of histories, making this FOM formulation largely independent of the history count. 
This specific FOM value enables comparisons across different VRT applications using the same code, hardware, and problem setup. 
Higher FOM values indicate greater efficiency, reflecting a method that achieves the desired accuracy with reduced computational cost.
The absolute magnitude of the FOM value is not particularly meaningful, just the relative value between methods.
For this reason, FOM results in the following tables have been normalized to 1.0 for a "base" case (typically the analog simulation).
This is not the case for the plots where the FOM magnitudes can vary.

The runtimes of a given simulation must include the computational cost of performing additional helper simulations.
The runtimes of the deterministic simulations, as well as the small-scale MC/DC simulations used to create the various weight window maps, are added to the final transport runtime.
MC/DC also exploits just-in-time (JIT) compilation, which needs to be factored into the runtime accounting.
Unlike compiled C++ codes like Shift (whose compilation times are never considered for these kinds of calculations), MC/DC compiles its Python kernels as needed during runtime.
While useful for MC/DC's other goals, this compilation time needs to be subtracted from the total runtime.
This can be accomplished by running MC/DC with only a few particles and caching the compiled kernels, which makes the subsequent simulations far more comparable to a pre-compiled code.
All runtimes reported in the following tables take into account these adjustments.

\subsection{Pulsed Slab}

\begin{figure}
  \centering
  \includegraphics[trim = 10mm 0mm 10mm 0mm, width=140mm]{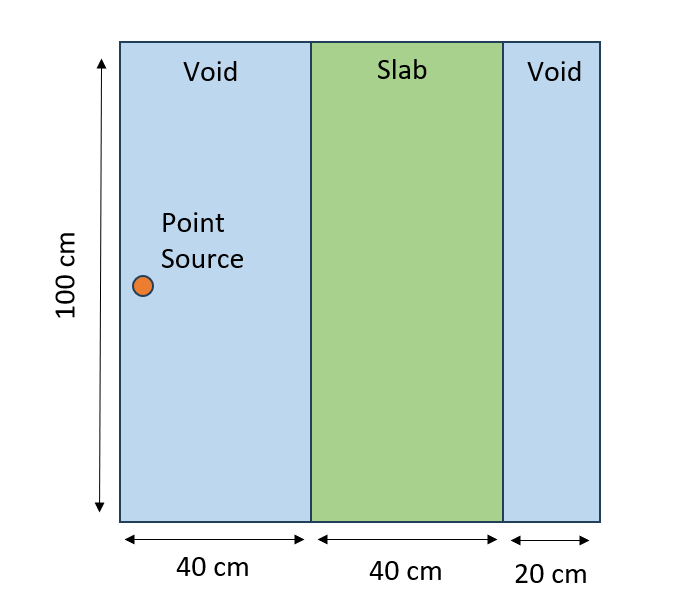}
  \caption{Geometry of the pulsed slab problem in Shift.}
  \label{PulsedSlabGeom}
\end{figure}

\begin{figure}
  \centering
  \includegraphics[trim = 10mm 0mm 10mm 0mm, width=140mm]{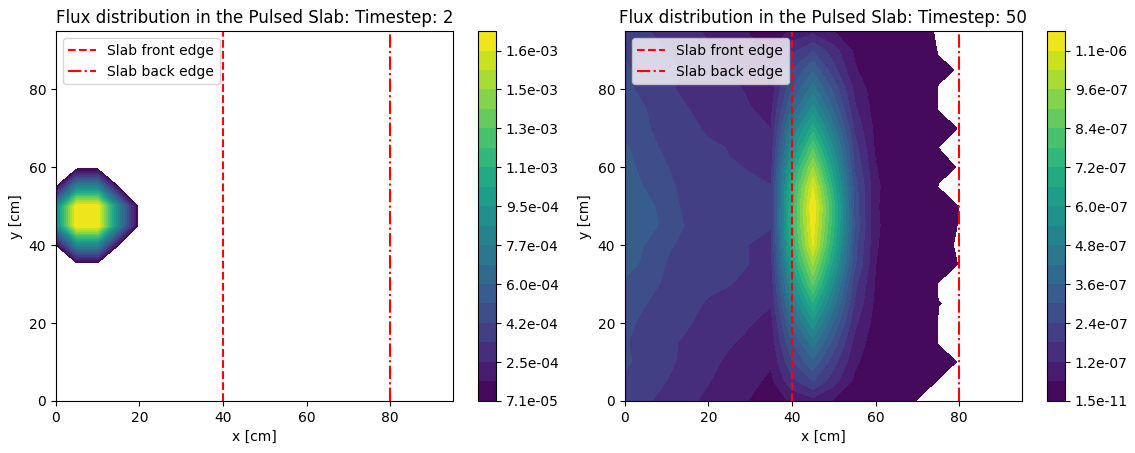}
  \caption{Flux distribution in an XY slice of the pulsed slab at the start and end of the simulation.}
  \label{PulsedSlabCombo}
\end{figure}

In Shift, a pulsed slab problem was developed to model neutron transport through the defined geometry pictured in Figure \ref{PulsedSlabGeom}. 
The problem consists of a box with dimensions of $100$ cm$^3$  and vacuum boundary conditions. 
The box is filled with $^{56}$Fe from $x = 40$ to $x = 80$ cm, while the remaining regions are void. 
Neutrons are emitted from a point source located at $x = 10$, $y = 50$, and $z = 50$ cm with an energy of 1 keV. 
Due to Shift’s current time-dependent source limitations, all source particles are initialized at $t = 0$. 
Over a sequence of 50 time steps, each is 100 ns in duration, $10^8$ source particles propagate from the point source and interact with the iron slab.
Delayed neutrons are not possible in Shift and would add an additional level of complexity to PCTs, but are not necessary to specifically investigate the interactions of PCTs with VCTs.
For the adjoint methods, a steady-state DENOVO solution was used to calculate the weight windows which could not change over time.
A 20x20x20 global tally mesh calculates the scalar flux and its variance to compare various algorithmic approaches.
The flux is plotted in Figure \ref{PulsedSlabCombo} for a timestep when the pulse is just beginning to move outwards and at the end of the simulation.

\begin{table}[htb]
  \centering
  \caption{Performance for the Pulsed Slab simulations in Shift.}
  \begin{tabular}{llllllllll}\toprule
    VRT & Runtime [s] & Average FOM
\\ \midrule
Analog & $104.1$ & $1.000$
\\
IC & 105.9 & $1.103$
\\
CADIS & $136.3$ & $0.740$
\\
FW-CADIS & $167.3$ & $1.248$
\\
\bottomrule
\end{tabular}
  \label{tab:pulsedslab}
\end{table}

Table \ref{tab:pulsedslab} displays the runtime and average figure of merit (FOM) across all tally bins in both space and time. 
For adjoint-based methods, the deterministic solves (each taking approximately two seconds) were included in the total runtime values.
Since this problem involves global tallies, FW-CADIS is expected to perform better than CADIS.
CADIS performs poorly in general when tasked with biasing particles towards a global mesh rather than a local tally since it encodes no information about how particles from the actual source reach areas of the problem.
FW-CADIS does perform the best out of all the tested VRTs, but the efficiency gains in the final Monte Carlo step only slightly outweigh the cost of the additional deterministic steps for this problem.
Challenges between Cooper-Larsen weight windows and weight-based combing could be affecting these simulations.
These challenges will be discussed in further detail with the following test problems.

\begin{figure}
  \centering
  \includegraphics[trim = 10mm 0mm 10mm 0mm, width=140mm]{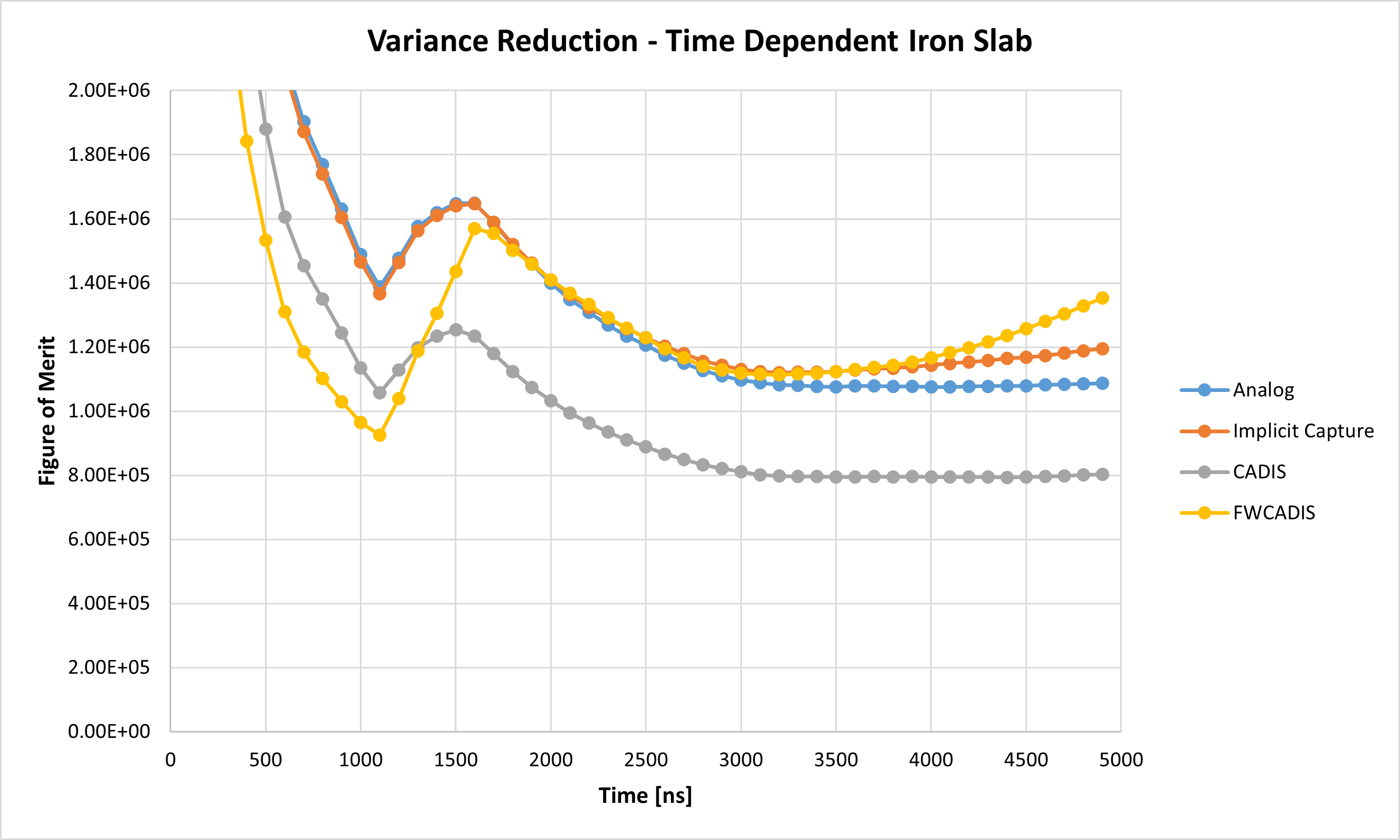}
  \caption{Spatially averaged FOM vs. time for the pulsed slab.}
  \label{PulsedSlabTD}
\end{figure}

An interesting result can be gleaned from Figure \ref{PulsedSlabTD}, which shows the spatially averaged FOM for the global mesh tally at each time step.
While the precise shape of the curve is not particularly relevant, the relative behavior of the methods over time is of interest.
The additional runtime burden of FW-CADIS reduces the method's FOM values at all time steps, but by the later time steps, the method catches up and begins to increase the rate at which it outperforms the other methods.
In the earlier time steps, the particles are mostly streaming through a void, rendering the VRTs ineffective.
As the particles penetrate the slab, the VRTs allow for more precise information in a greater range of the spatial domain.
Simulations in which the flux distribution becomes more spatially heterogeneous over time might benefit more from the use of FW-CADIS.

Multiple factors likely contribute to the modest efficiency improvements observed in the use of FW-CADIS. 
First, the weight windows lack time dependence, hence, the same window mesh is applied at every time step despite considerable changes in neutron flux distributions over the course of the simulation. 
Second, weight-based combing is consistently active, which may cause the weight windows to interfere with the regulation of particle weights from the combing. 
Lastly, this problem's simplicity and small physical scale may naturally limit the need for extensive variance reduction to achieve efficient results. 
Despite these constraints, the unoptimized use of existing global VRTs still demonstrates advantages in time-dependent contexts.

\subsection{AZURV1}

The AZURV1 benchmarks~\cite{Ganapol2001} were developed to provide a set of analytical standards for time-dependent neutron transport problems. 
This analysis focuses on a supercritical, one-dimensional infinite-medium point source problem from the set.
To model this scenario, an infinite one-dimensional plane was set up in CSG by defining two parallel planes with reflecting boundary conditions, creating a confined space filled with a homogeneous material that has simplified cross-section data. 
Capture, scattering, and fission each have a macroscopic cross section of one-third $cm^{-1}$, with the average number of neutrons released per fission set to 2.3.
Neutrons from both scattering and fission events are emitted isotropically.

This configuration produces a multiplication factor of 1.15, resulting in a supercritical system and a neutron population that increases over time. 
Running this simulation without some form of population control is infeasible because the particle population will exponentially increase, rapidly using up available memory.

The neutrons travel at one centimeter per second, and the simulation terminates after 20 time steps of length one second. 
At $x=0$ and $t=0$, an isotropic neutron point source initiates a pulse propagating through the x-dimension. 
A tally mesh consisting of 201 uniform bins covers the x-domain, ranging from -20.5 to 20.5 cm while 20 time bins capture neutron behavior from 0 to 20 seconds.
Due to the significant (orders of magnitude) decrease in neutron flux at the periphery of the pulse, variance reduction can dramatically affect the efficiency over sections of this problem.
During the 20-second simulation period, the neutrons will travel a maximum of 20 cm in either the positive or negative direction, however, analog Monte Carlo simulations typically fail to transport the vast majority of particles beyond approximately 15 cm.

Analog simulations, IC, and CL weight windows were simulated with both UC and WC enabled.
The CL weight windows were produced from an initial small-scale MC/DC simulation using a small fraction of the particles (1E4 particles instead of the 1E6 used for the full scale simulation).
The flux from this small-scale simulation is tallied, normalized, and imported into the full-scale MC/DC simulation as weight window centers.
In locations where the flux is zero, the weight windows are adjusted to the smallest nonzero value in the weight window mesh, as zero-valued window centers lead to mathematical complications.
The computational cost of the small scale simulation must be accounted for in the FOM calculation.
In addition to the standard spatial mesh for CL weight windows, supplementary sets of CL simulations were conducted employing 2D spatial and time-dependent meshes.
The time bins on the weight window mesh were set equal to the time steps of the simulation which means that a new spatial weight window mesh is used at each time step to adjust to the changing flux distribution.

\begin{table}[htb]
  \centering
  \caption{Statistics for the AZURV1 benchmark in MC/DC.}
  \begin{tabular}{llllllllll}\toprule
    VRT & PCT & Runtime [s] & Average FOM
\\ \midrule
Analog & UC & $414.6$ & $1.00$
\\
Analog & WC & $408.6$ & $1.02$
\\
IC & UC & $419.6$ & $7.83E-4$
\\
IC & WC & $415.2$ & $0.577$
\\
CL (static weight windows) & UC & $418.2$ & $1.087E+4$ 
\\
CL (dynamic weight windows)& UC & $416.8$ & $2.087E+5$
\\
\bottomrule
\end{tabular}
  \label{tab:azurv1}
\end{table}

MC/DC's performance on the AZURV1 problem is summarized in table \ref{tab:azurv1}.
A valuable sanity check is the similarity in FOM between UC and WC in an analog simulation.
When all particles possess the same weight, these PCT methods are essentially identical.
A major point to note is the lack of data on CL weight windows in conjunction with WC. This stems from the fact that every simulation with this combination exhausted memory prior to completion, despite the allocation of extremely large overflow particle banks.
The main cause for this conflict is the resetting of particle weights in WC.
Significant splitting occurs near the pulse edge due to the weight windows attempting to adjust for the large neutron flux gradient.
When WC resets particle weights at census, the weights of particles sourced into the next time step are no longer aligned with the weight window distribution.
This leads to excessive splitting at the pulse edge, rapidly depleting all available memory.
This problem is exacerbated in MC/DC due to the inability to normalize weight windows properly when WC is applied.
In UC, surviving particles have their weights multiplied by a constant factor which can also be used to renormalize the weight windows for the following time step, ensuring that particles are sourced into reasonable weight window ranges.
However, there is currently no analog to that rescale value when the particle weights get reset in WC.
This results in a disconnect between the weights of the particles and the normalization of the weight windows.

\begin{figure}
  \centering
  \includegraphics[trim = 10mm 0mm 10mm 0mm, width=140mm]{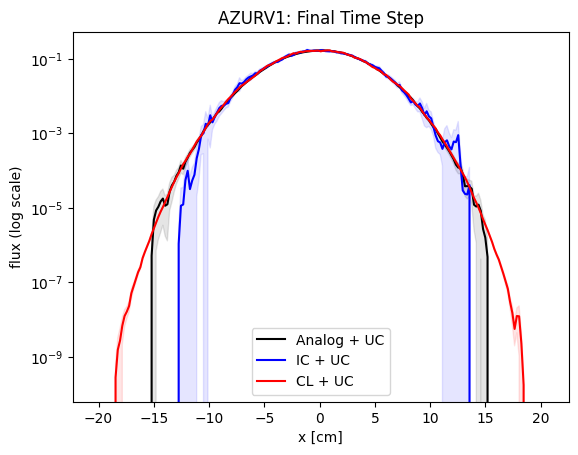}
  \caption{Neutron flux at the final time step of the AZURV1 simulations with error bars for two standard deviations.}
  \label{fig:azurv1}
\end{figure}

\begin{figure}
  \centering
  \includegraphics[trim = 10mm 0mm 10mm 0mm, width=140mm]{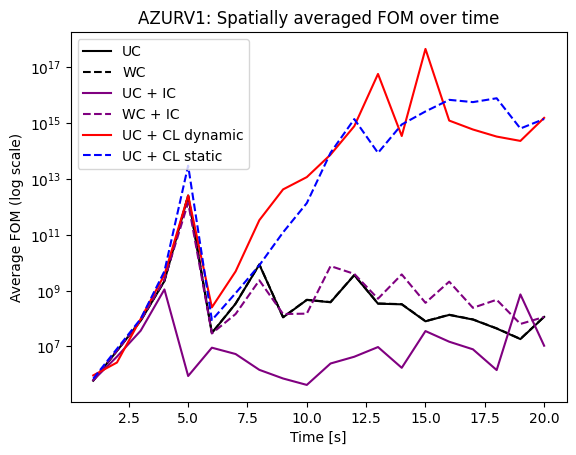}
  \caption{Spatially averaged FOM at each time step for the AZURV1 problem.}
  \label{fig:azurv1fomvtime}
\end{figure}

IC does not improve efficiency in this version of the AZURV1 problem in MC/DC, and unexpectedly exhibits weak synergy with UC.
This conflict appears in all test problems simulated so far in MC/DC using these methods.
To ensure this behavior isn't exclusively associated with a supercritical multiplying problem, the average neutrons released per fission was reduced from 2.3 to 1.7 to make the problem subcritical.
Results from a set of investigations into IC mixed with the combing methods are shown in Table \ref{tab:azurv1_sub}.
In this table a slightly different FOM was used, using the relative error and a harmonic mean instead of the pure standard deviation and spatial average, as compared to the supercritical table.
However, the important information in FOM values is just the \textit{relative} behavior between methods, and the subcritical FOM values follow the same pattern as those from the supercritical problem.
For this problem and method combination in particular, the average FOM is hampered by a significant lack of particles near the wavefront.
The initial hypothesis going into these simulations was that IC would coexist well with the UC method, which should not affect the distribution of particles and weights.

\begin{table}[htb]
  \centering
  \caption{Statistics for the subcritical AZURV1 benchmark in MC/DC.}
  \begin{tabular}{llllllllll}\toprule
    VRT & PCT & Runtime [s] & Average FOM
\\ \midrule
Analog & UC & $409.8$ & $1.00$
\\
Analog & WC & $409.8$ & $1.00$
\\
IC & UC & $434.4$ & $0.767$
\\
IC & WC & $436.8$ & $1.152$
\\
\bottomrule
\end{tabular}
  \label{tab:azurv1_sub}
\end{table}

\begin{figure}
  \centering
  \includegraphics[trim = 10mm 0mm 10mm 0mm, width=140mm]{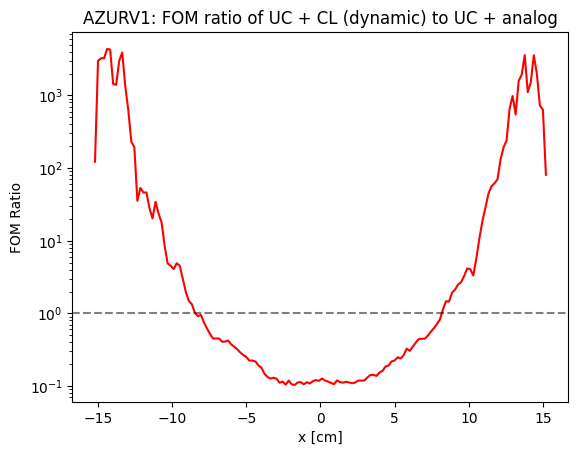}
  \caption{Spatial distribution of the FOM ratio for CL weight windows vs. analog at the final time step of the AZURV1 problem.}
  \label{fig:azurv1spatialfomratio}
\end{figure}

Some key insights from Table \ref{tab:azurv1} lie in the significantly improved FOM achieved by using Cooper-Larsen (CL) weight windows in combination with uniform combing (UC). 
This improvement is evident in Figure \ref{fig:azurv1}, which illustrates that the analog simulation struggles to produce reliable tallies at the boundaries of the problem, yielding good statistics only near the center. 
Figure \ref{fig:azurv1fomvtime} also shows that this benefit of CL weight windows is more prevalent at later time steps when the particles have had a chance to spread further.
At the edges, where the flux is low, the CL weight windows employ frequent particle splitting, which allows for better statistical results in those regions.
This can clearly be seen in figure \ref{fig:azurv1spatialfomratio}.
The CL weight windows also roulette frequently towards the center of the problem, decreasing the FOM there, but this is acceptable because the center is already statistically over-resolved.

Additionally, the time-dependent weight windows outperformed their steady-state counterparts. 
By adapting to the changing neutron flux over time, they achieved a modest increase in efficiency. 
This serves as a solid initial demonstration of how steady-state methods can be adapted for time-dependent simulations. 
As more simulations integrate time-dependence, optimizing variance reduction techniques for this extra dimension may continue to provide significant efficiency gains ~\cite{Shaw2025, Legrady2020}.

\subsection{1D SCRAM}

Another test problem leverages some of MC/DC's time-dependent modeling capabilities, particularly surfaces that move continuously with time \cite{Variansyah2023}.
This energy-independent problem involves a one-dimensional slab that extends from zero to six centimeters along the x-axis, with vacuum boundary conditions on either end. 
A surface divides the slab into two distinct regions: a scattering medium on the lower end of x and an absorbing medium on the upper end. 
The scattering medium has a macroscopic scattering cross section of 0.9 $cm^{-1}$ and a capture cross section of 0.1 $cm^{-1}$. 
In contrast, the absorbing medium has a scattering cross section of 0.1 $cm^{-1}$ and a capture cross section of 0.9 $cm^{-1}$.

\begin{figure}
  \centering
  \includegraphics[trim = 10mm 0mm 10mm 0mm, width=140mm]{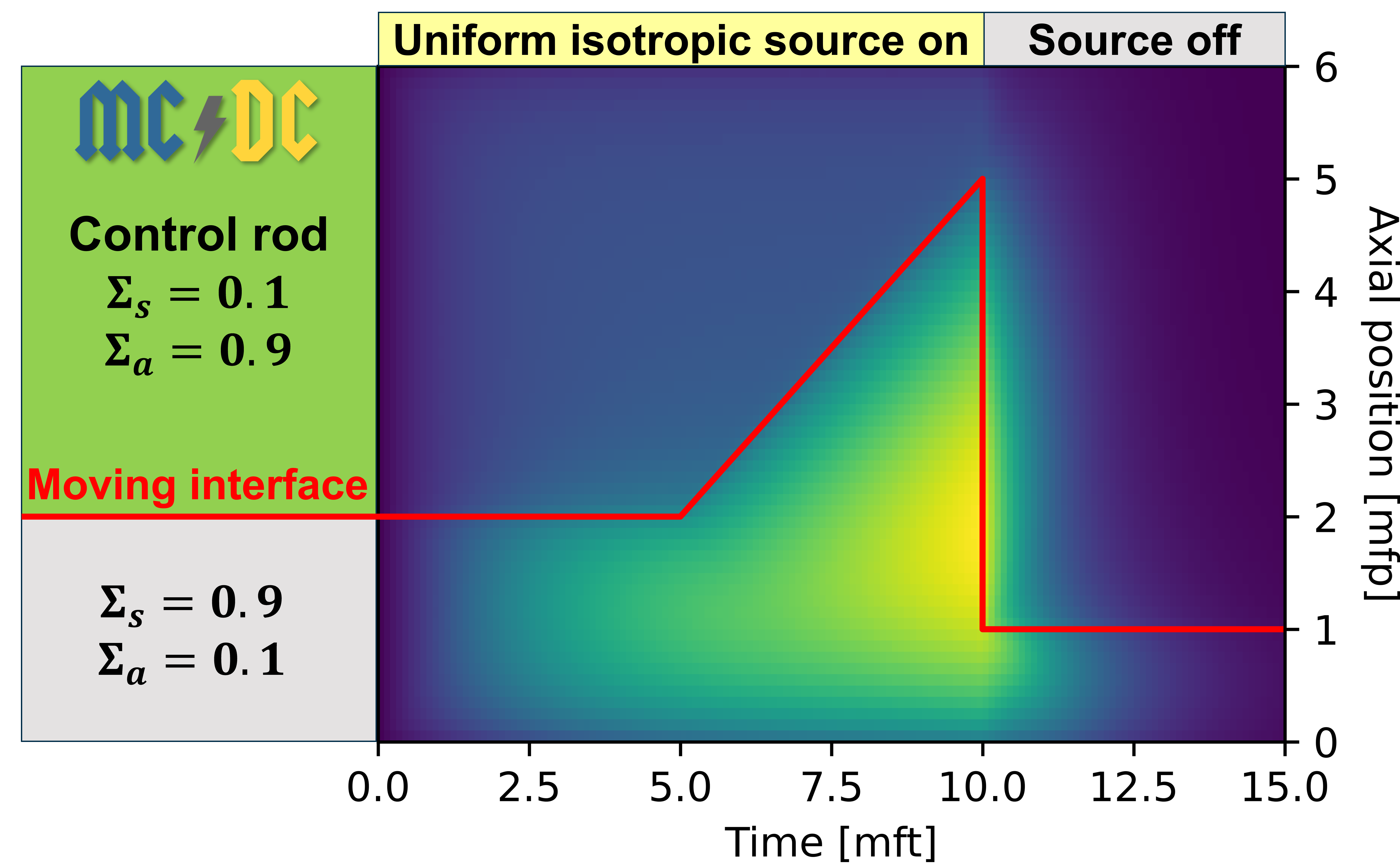}
  \caption{Flux profile for the 1D SCRAM problem \cite{Variansyah2023}}
  \label{1DSCRAMflux}
\end{figure}

\begin{figure}
  \centering
  \includegraphics[trim = 10mm 0mm 10mm 0mm, width=140mm]{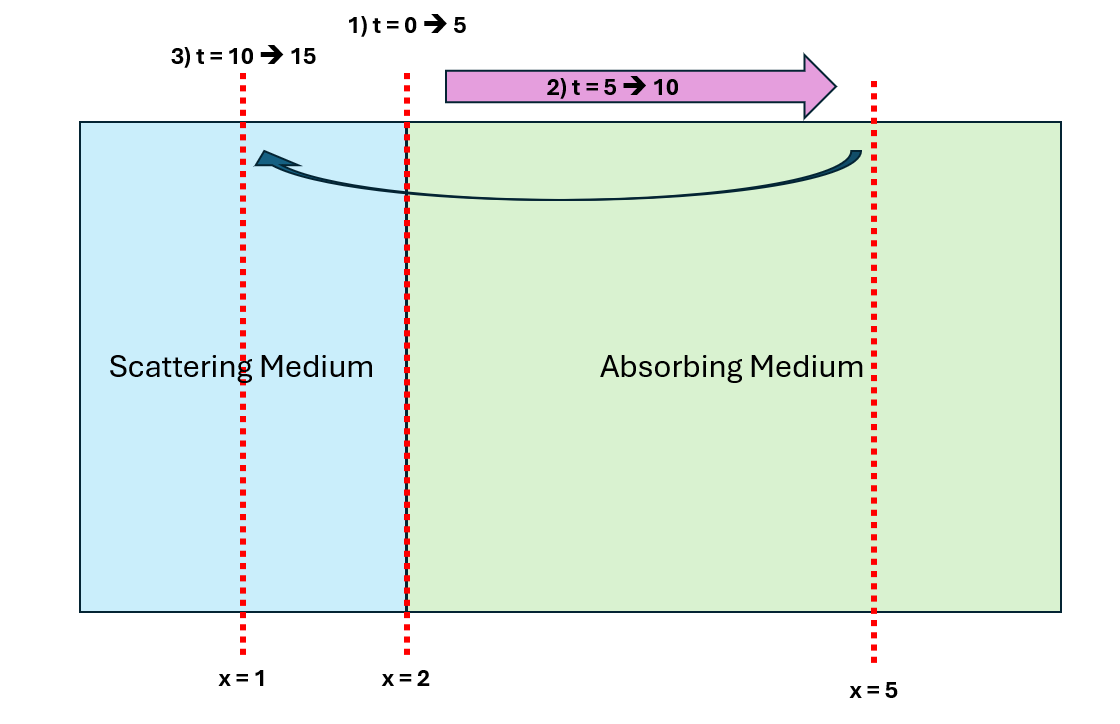}
  \caption{Visual rendition of the 1D reactor SCRAM simulation setup.}
  \label{1DSCRAMGeom}
\end{figure}

The dividing surface is initially positioned at $x=2$ $cm$ and remains there for the first five seconds. 
From $t=5$$s$ to $t=10s$, it shifts linearly and continuously from $x=2$ $cm$ to $x=5$ $cm$. 
At $t=10s$, the surface instantaneously moves down to $x=1$ $cm$, where it stays for the remainder of the simulation.
Particles are uniformly and isotropically emitted across the entire domain for the first 10 seconds with a speed of 1 cm/s. 
The neutron flux grows in the scattering region as it expands, then dies off when the absorbing region pushes back in as shown in Figure \ref{1DSCRAMflux}.
This movement, shown in Figure \ref{1DSCRAMGeom}, is a simplistic rendition of a control rod being slowly withdrawn and then rapidly inserted into a reactor.

The simulation uses time steps of 0.5 seconds, totaling a duration of 15 seconds and 30 time-steps with census and population control. 
A global tally mesh segments the x-dimension into 60 evenly spaced bins and time into 30 uniform intervals.
Analog simulations, IC, and CL weight windows were all used with no PCT, UC, and WC, and the simulation results are shown in Table \ref{tab:1dSCRAM}.
In that table, the spatially and temporally averaged FOM values are reported.
CL weight windows without PCT results in poor management of the particle population, and to keep the runtimes reasonable, $10^6$ particles were used as opposed to the $5*10^6$ for all the other method combinations.
The FOM comparison is still valid since it has been chosen to be independent of the number of initial particles.

\begin{table}[htb]
  \centering
  \caption{Statistics for the 1D SCRAM problem in MC/DC.}
  \begin{tabular}{llllllllll}\toprule
        Variance & Population & Runtime [s] & Average \\
    Reduction &  Control & & FOM \\
\\ \midrule
Analog & None &  105.6 & 1.00\\
\\
IC & None  & 166.2 & 0.980\\
\\
CL & None & 620.4* & 0.761\\
\\
Analog & UC & 169.2 & 34.0\\
\\
IC & UC & 202.2 & 1.32\\
\\
CL & UC & 204.2 & 49.6\\
\\
Analog & WC & 169.8 & 31.7\\
\\
IC & WC & 171.6 & 34.5\\
\\
\bottomrule
\end{tabular}
  \label{tab:1dSCRAM}
\end{table}

\begin{figure}
  \centering
  \includegraphics[trim = 10mm 0mm 10mm 0mm, width=140mm]{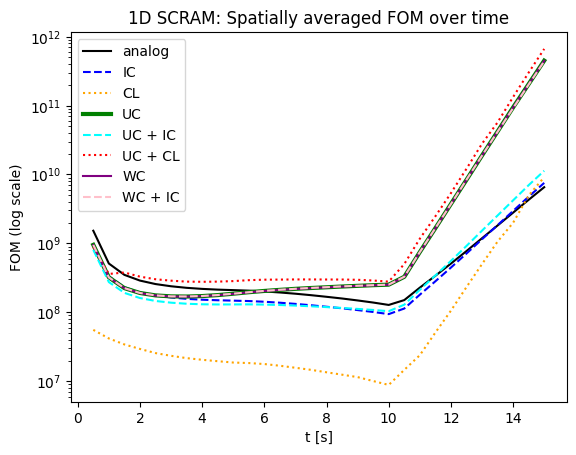}
  \caption{Spatially averaged FOM at each time step for the 1D SCRAM problem.}
  \label{1DSCRAMFOMvtime}
\end{figure}

Figure \ref{1DSCRAMFOMvtime} displays the spatially averaged FOM values at each time step.
The advantage of the best performing methods such as UC + CL weight windows is particularly noticable when the absorbing material is "inserted" in the ending time steps.
The comparatively larger differences in flux magnitudes across the problem during those steps allow for the VRTs to have a greater impact on particle/weight distributions.

This problem demonstrates the value of applying a PCT in time-dependent simulations.
Without explicit controls over the number of particles in the system, managing the memory and runtimes of simulations becomes much more difficult.
The overall efficiency of the simulation decreases in the absence of a PCT because some time steps have far fewer particles than others.
Implementing fully time-dependent weight windows could help address this issue by acting as a loose PCT themselves.
Such methods have been used in other codes, and are currently being developed in MC/DC ~\cite{Shaw2025}.

The results with PCTs mirror the AZURV1 benchmark results with less drastic differences between method combinations.
Once more, implicit capture significantly underperformed when paired with uniform combing, while the combination of CL weight windows and uniform combing performed the best.
Again, weight-based combing combined with CL weight windows did not function.
The same splitting issues were observed, leading to overloaded particle banks in memory.

\section{Conclusion}
\label{sec:conclusion}

Selecting appropriate population control and variance reduction algorithms is crucial for time-dependent Monte Carlo radiation transport simulations.
Using unsuitable combinations of methods can damage the simulation's efficiency and effectiveness.
In particular, this work highlights the adverse impacts of combining weight windows with weight based combing, and implicit capture with uniform combing.

Conversely, selecting the proper methods for the problem can lead to significantly more efficient simulations, possibly making computationally expensive, time-dependent analysis viable.
Cooper-Larsen weight windows offer valuable global variance reduction when combined with uniform combing.
More sophisticated weight windows techniques like the CADIS methods \cite{Wagner1998, Wagner2014} or MAGIC \cite{Davis2011} may interact favorably with uniform combing and other population control techniques, which do not disturb the distributions of particles and weights across the problem domain.

A subset of variance reduction techniques originally developed for steady-state simulations have been shown to be effective when applied to time-dependent problems.
Notably, time-dependent adaptations of these techniques yield promising results for further optimizations.
The ability of these methods to scale to more complex time-dependent problems utilizing massively parallel computing architectures will be vital for advancing future simulations.
Additional existing variance reduction \cite{Davis2011} and population control techniques \cite{Booth1996}, and their steady-state or potential time-dependent variations,  could greatly assist in simulating a wider range of dynamic problems that are just becoming achievable.

\pagebreak
\section*{Acknowledgments}

This work was supported by the Center for Exascale Monte-Carlo Neutron Transport
(CEMeNT) a PSAAP-III project funded by the Department of Energy, grant number
DE-NA003967.

In accordance with publishing guidelines, it is hereby disclosed that ChatGPT (versions 3.5 and 4o) was used for minor text revisions in this document.

The authors report there are no competing interests to declare.

\pagebreak
\bibliographystyle{style/ans_js}                                                                           
\bibliography{bibliography}

\end{document}